\documentclass[a4paper,lnbip]{svmultln}
\usepackage{graphicx}
\usepackage{latexsym}
\usepackage{float}
\usepackage{amssymb}
\usepackage{amsmath}
\usepackage{algorithm}
\usepackage[noend]{algorithmic}
\usepackage{tabularx}

\graphicspath{{img/}}

\begin{document}
\mainmatter 

\title{Expressiveness and Understandability Considerations of Hierarchy in
Declarative Business Process Models\thanks{This research is supported by Austrian
Science Fund (FWF): P23699-N23 and the BIT fellowship program. The final
publication is available at Springer via
http://dx.doi.org/10.1007/978-3-642-31072-0\_12}}
\titlerunning{Considerations of Hierarchy in Declarative Process Models}
\author{Stefan Zugal\inst{1} \and Pnina Soffer\inst{2} \and Jakob
Pinggera\inst{1} \and Barbara Weber\inst{1}} \authorrunning{Stefan Zugal \and Pnina Soffer \and Jakob
Pinggera \and  Barbara Weber}

\institute{University of Innsbruck, Austria\\
\email{{stefan.zugal, jakob.pinggera, barbara.weber}@uibk.ac.at} \and
University of Haifa, Israel\\
\email{spnina@is.haifa.ac.il}
}
\maketitle

\begin{abstract}
Hierarchy has widely been recognized as a viable approach to deal with the
complexity of conceptual models. For instance, in declarative business process
models, hierarchy is realized by sub-processes. While technical implementations
of declarative sub-processes exist, their application, semantics, and the
resulting impact on understandability are less understood yet---this research gap
is addressed in this work. In particular, we discuss the semantics and the
application of hierarchy and show how sub-processes enhance the expressiveness of
declarative modeling languages. Then, we turn to the impact on the
understandability of hierarchy on a declarative process model. To systematically
assess this impact, we present a cognitive-psychology based framework that allows
to assess the possible impact of hierarchy on the understandability of the
process model.
\end{abstract}

\section{Introduction}
Using modularization to hierarchically structure information has for decades been
identified as a viable approach to deal with complexity~\cite{Parn72}. Not
surprisingly, business process modeling languages provide support for
hierarchical structures, e.g., sub-processes in BPMN and YAWL. However, in
general, \textit{``the world does not represent itself to us neatly divided into
systems, subsystems\ldots these divisions which we make
ourselves''}~\cite{GoVa79}. In this sense, a viable discussion about the proper
use of modularization for the analysis and design of information
systems as well as its impact on understandability is still going
on. In business process management, sub-processes have been recognized as an
important factor influencing model understandability~\cite{Dami07},
however, there are no definitive guidelines on their use yet. For instance,
recommendations regarding the size of a sub-process in an \textit{imperative
process model} range from 5--7 model elements~\cite{ShMc01} over 5--15 model
elements~\cite{Kock96} to up to 50 model elements~\cite{MeRA10}. For declarative
process models, which have recently gained attention due to their
flexibility~\cite{Pes+07}, the proper usage of modularization is even less clear.
While work has been done with respect to the technical support of declarative
sub-processes, it remains unclear \textit{whether and when} hierarchy has an
influence on the \textit{understandability} of the model. In general, empirical
research into the understandability of conceptual models (e.g., ER diagrams or
UML statecharts) has shown that hierarchy can have a positive
influence~\cite{Rej+11}, negative influence~\cite{Cru+08} or no influence at
all~\cite{Cru+05a}. For declarative process models, the situation is less clear as
no empirical studies have been conducted so far. However, as declarative process
models appear to be especially challenging to understand~\cite{ZuPWxx}, it seems
particularly important to improve their understandability. In the following, we
will shed light on the question which influence on understandability can be
expected for hierarchy in declarative process models.

The contribution of this work is twofold. First, the semantics of hierarchy in
declarative process models is elaborated on. In particular, we will show that
hierarchy is not just a question of structure, but also enhances expressiveness
and has implications on the restructuring of a model. Second, the impact of
hierarchy on the understandability of the model will be investigated
systematically. We will present a cognitive-psychology based framework that
explains general effects of hierarchy, but also takes into account peculiarities of
declarative process models. The framework allows to assess the possible impact of
hierarchy, i.e., whether a certain modularization of a declarative process model
has a positive influence, negative influence or no influence at all. This, in
turn, allows for a systematic empirical investigation in future
work.\footnote{Please note that even though we take into account declarative
models in general, we will make use of the declarative language
\textit{ConDec}~\cite{Pesi08} for the discussion.}

The remainder of this paper is structured as follows.
Section~\ref{sec:declarative_processes} introduces declarative process models.
Then, Section~\ref{sec:semantics} discusses the semantics of hierarchy in
declarative process models. Subsequently, Section~\ref{sec:using_hierarchy} deals
with the application of hierarchy in declarative process models, whereas
Section~\ref{sec:understandability} investigates the impact on understandability.
Finally, related work is presented in Section~\ref{sec:related_work} and the
paper is concluded with a summary and an outlook in Section~\ref{sec:summary}.

\section{Background: Declarative Processes}
\label{sec:declarative_processes}
There has been a long tradition of modeling business processes in an imperative
way. Process modeling languages supporting this paradigm, like BPMN, EPC and UML
Activity Diagrams, are widely used. Recently, \textit{declarative approaches}
have received increasing interest and suggest a fundamentally different way of
describing business processes~\cite{Pesi08}. While imperative models specify
exactly \textit{how} things have to be done, declarative approaches only focus on
the logic that governs the interplay of actions in the process by describing the
\textit{activities} that can be performed, as well as \textit{constraints}
prohibiting undesired behavior. An example of a constraint in an aviation process
would be that crew duty times cannot exceed a predefined threshold. Constraints
described in literature can be classified as execution and termination
constraints. \textit{Execution} constraints, on the one hand, restrict the
execution of activities, e.g., an activity can be executed at most once.
\textit{Termination} constraints, on the other hand, affect the \textit{proper}
termination\footnote{In the following we will use \textit{termination} as synonym
for \textit{proper termination}.} of process instances and specify when process
termination is possible. For instance, an activity must be executed at least once
before the process can be terminated. Most constraints focus either on execution
\textit{or} termination semantics, however, some constraints also combine
execution and termination semantics (e.g., the succession
constraint~\cite{Pesi08}).

\begin{figure}[htp]
 \centering
 \includegraphics[width=.90\textwidth]{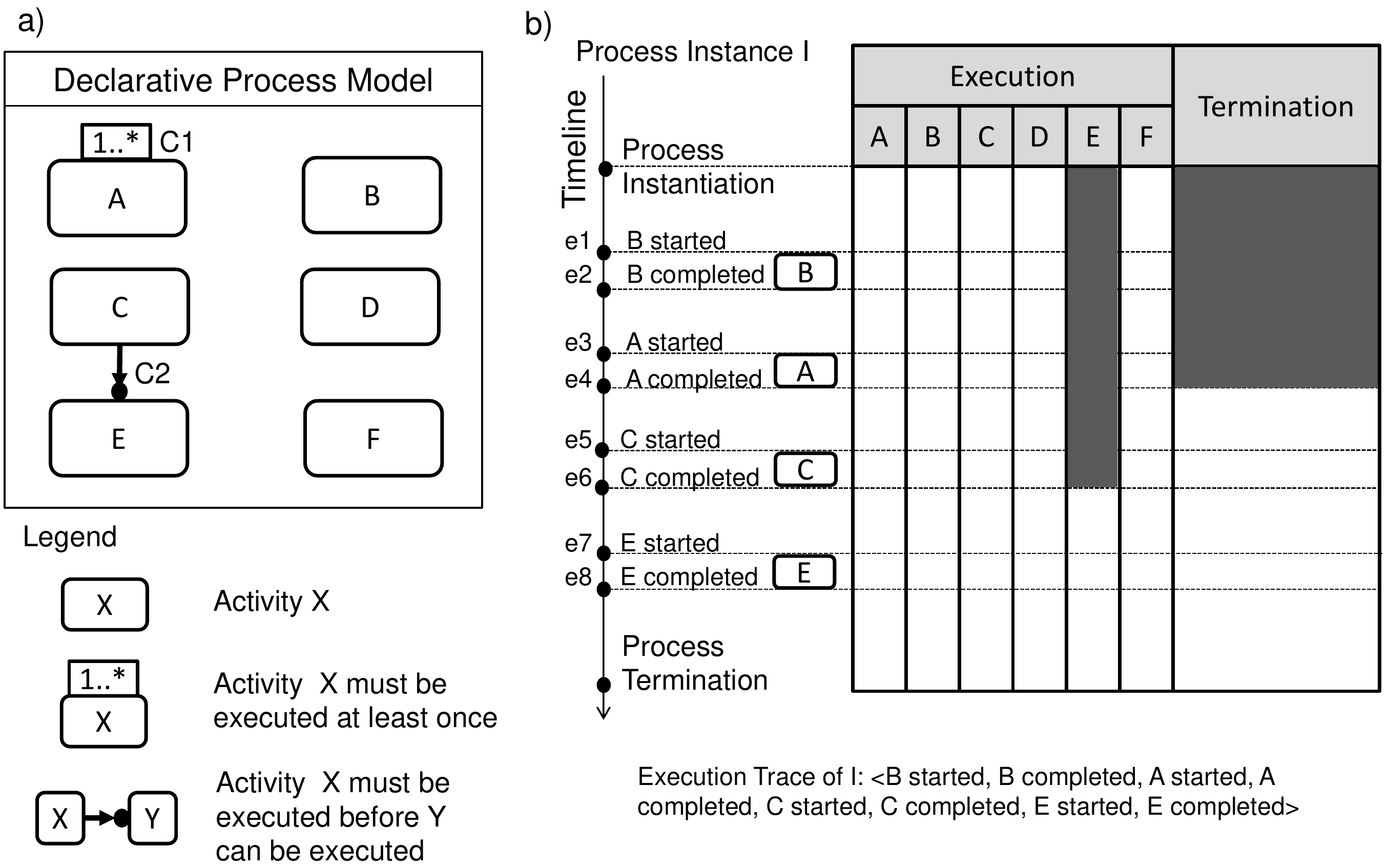}
 \caption{Executing a declarative process}
 \label{fig:execution_termination_semantics}
\end{figure}

To illustrate the concept of declarative processes, a model specified in
ConDec~\cite{Pesi08} is shown in
\figurename~\ref{fig:execution_termination_semantics}~a). It contains activities
\textit{A} to \textit{F} as well as constraints \textit{C1} and \textit{C2}.
\textit{C1} prescribes that A must be executed at least once (i.e.,
\textit{C1} restricts the termination of process instances). \textit{C2}
specifies that \textit{E} can only be executed if \textit{C} has been executed at
some point in time before (i.e., \textit{C2} imposes restrictions on the
execution of activity E). In
\figurename~\ref{fig:execution_termination_semantics}~b) an example of a process
instance illustrates the semantics of the described constraints. After process
instantiation, \textit{A}, \textit{B}, \textit{C}, \textit{D} and \textit{F} can
be executed. \textit{E}, however, cannot be executed as \textit{C2} specifies
that \textit{C} must have been executed before (cf. grey bar in
\figurename~\ref{fig:execution_termination_semantics}~b) below ``E'').
Furthermore, the process instance cannot be terminated as \textit{C1} is not
satisfied, i.e., \textit{A} has not been executed at least once (cf. grey area in
\figurename~\ref{fig:execution_termination_semantics}~b) below ``Termination'').
The subsequent execution of B does not cause any changes as it is not involved in
any constraint. However, after \textit{A} is executed, \textit{C1} is satisfied,
i.e., \textit{A} has been executed at least once and thus the process instance
can be terminated (cf.
\figurename~\ref{fig:execution_termination_semantics}~b)---after \textit{e4} the
box below ``Termination'' is white). Then, \textit{C} is executed, satisfying
\textit{C2} and consequently allowing \textit{E} to be executed (the box below
``E'' is white after \textit{e6} occurred). Finally, the execution of \textit{E}
does not affect any constraint, thus no changes with respect to constraint
satisfaction can be observed. As all
 termination constraints are still satisfied, the process instance can still be
 terminated. Please note that declarative process instances have to be terminated
 explicitly, i.e., the end user must decide when to complete the process
 instance. Termination constraints thereby specify when termination is allowed,
 i.e., the process instance I could have been terminated at any point in time
 after \textit{e4}.

As illustrated in \figurename~\ref{fig:execution_termination_semantics}, a
process instance can be specified through a list of \textit{events} that
describe changes in the life-cycle of \textit{activity instances}, e.g.,
\textit{``e1: B started''}. In the following, we will denote this list as
\textit{execution trace}, e.g., for process instance I: $<$\textit{e1},
\textit{e2}, \textit{e3}, \textit{e4}, \textit{e5}, \textit{e6}, \textit{e7},
\textit{e8}$>$. If activities are non-overlapping, we merge subsequent start
events and completion events, e.g., $<$\textit{B started}, \textit{B completed},
\textit{A started}, \textit{A completed}$>$ is abbreviated by $<$\textit{B},
\textit{A}$>$.

\section{Discussion of Semantics}
\label{sec:semantics}
This section aims at establishing an understanding of the semantics of
sub-processes in a declarative model. Based on this, the next section discusses
their possible use, and then we turn to discuss their possible effect on model
understanding. To our knowledge, the semantics of declarative sub-processes have
not been discussed explicitly yet, but their use has been suggested in the
context of imperative-declarative model combinations~\cite{Pes+07b}. In general,
a sub-process is introduced in a process model via a \textit{complex activity},
which refers to a process model. When the complex activity is executed, the
referred process model, i.e., the sub-process, is instantiated. Thereby,
sub-processes are viewed as \textit{individual} process instances, i.e., when a
complex activity is started, a new instance of the sub-process the complex
activity is referring to is created (cf.~\cite{Pes+07b}). The parent
process, however, has no information about the internals of the sub-process,
i.e., the sub-process is executed in isolation. Communication with the parent
process is done only via the sub-process' life-cycle\footnote{We do not take into
account communication via input- and output data here, as we focus on control
flow behavior only.}. Thereby, the life-cycle state of the complex activity
reflects the state of the sub-process~\cite{Pes+07b}, e.g., when the complex
activity is in state \textit{completed}, also the sub-process must be in state
\textit{completed}.

Considering this, it is essential that sub-processes are executed in isolation,
as isolation forbids that constraints can be specified between activities
included in different sub-processes. In other words, in a hierarchical
declarative process model with several layers of hierarchy, the constraints of a
process model \textit{can neither directly} influence the control flow of any
parent process, \textit{nor directly influence} the control flow of any
sub-process on a layer below. Please note that control flow may still be
\textit{indirectly influenced} by restricting the execution of a sub-process,
thereby restricting the execution of the activities contained therein.

\begin{figure}[htp]
\begin{center}
  \includegraphics[width=\textwidth]{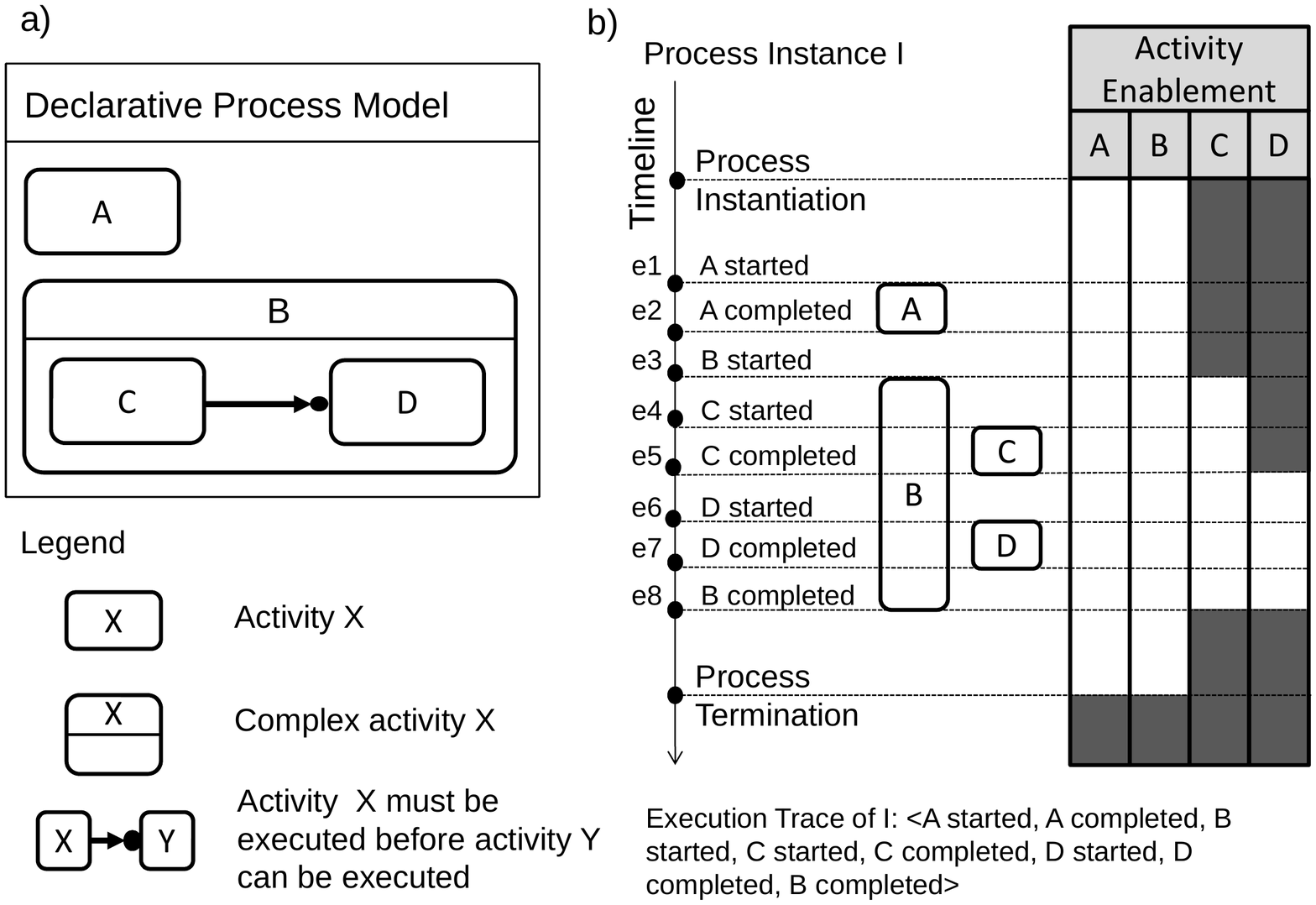}
  \caption{Execution of a sub-process}
  \label{fig:semantics}
\end{center}
\end{figure}

To illustrate these concepts, consider the process model in
\figurename~\ref{fig:semantics} a). It consists of activity \textit{A} and
complex activity \textit{B}, which in turn contains activities \textit{C} and
\textit{D}. \textit{C} and \textit{D} are connected by a precedence constraint,
i.e., \textit{D} can only be executed if \textit{C} was executed before.
\figurename~\ref{fig:semantics} b) shows an example of a process instance that is
executed on this process model. On the left a timeline lists all events that
occur during the process execution, e.g., starting or completing an activity. To
the right, the enablement of the activities is illustrated. Whenever the area
below an activity is colored white, it indicates that this activity is currently
enabled. The timeline is to be interpreted the following way: By instantiating
the process, activities \textit{A} and \textit{B} become enabled, as no
constraints restrict their execution. \textit{C} and \textit{D} cannot be
executed, as they are confined in complex activity \textit{B} and no instance of
\textit{B} is running yet. Also the subsequent execution of \textit{A}
(\textit{e1}, \textit{e2}) does not change activity enablement. However, with the
start of \textit{B} (\textit{c3}), \textit{C} becomes enabled, as it can be
executed within the new instance of \textit{B}. Still, \textit{D} is not enabled
yet as the precedence constraint is not satisfied. After \textit{C} is executed
(\textit{e4}, \textit{e5}), the precedence constraint is satisfied, therefore
also \textit{D} becomes enabled. After the execution of \textit{D} (\textit{e6},
\textit{e7}), the user decides to complete sub-process \textit{B}
(\textit{e8}). Hence, \textit{C} and \textit{D} cannot be executed anymore.
Still, \textit{A} and \textit{B} are enabled as they can be executed directly
within process instance \textit{I}. Finally, after the process instance is
completed by the end user through explicit termination, no activity is enabled
anymore.

\section{Using Hierarchy in Declarative Models}
\label{sec:using_hierarchy}
Regardless of the modeling language, hierarchy allows to structure models and to
hide modeling elements in sub-models. In this section, the use of hierarchy,
given the semantics of Section~\ref{sec:semantics}, is discussed.

\subsection{Running Example}
\label{sec:example}
\begin{figure}[htp]
\begin{center}
  \includegraphics[width=\textwidth]{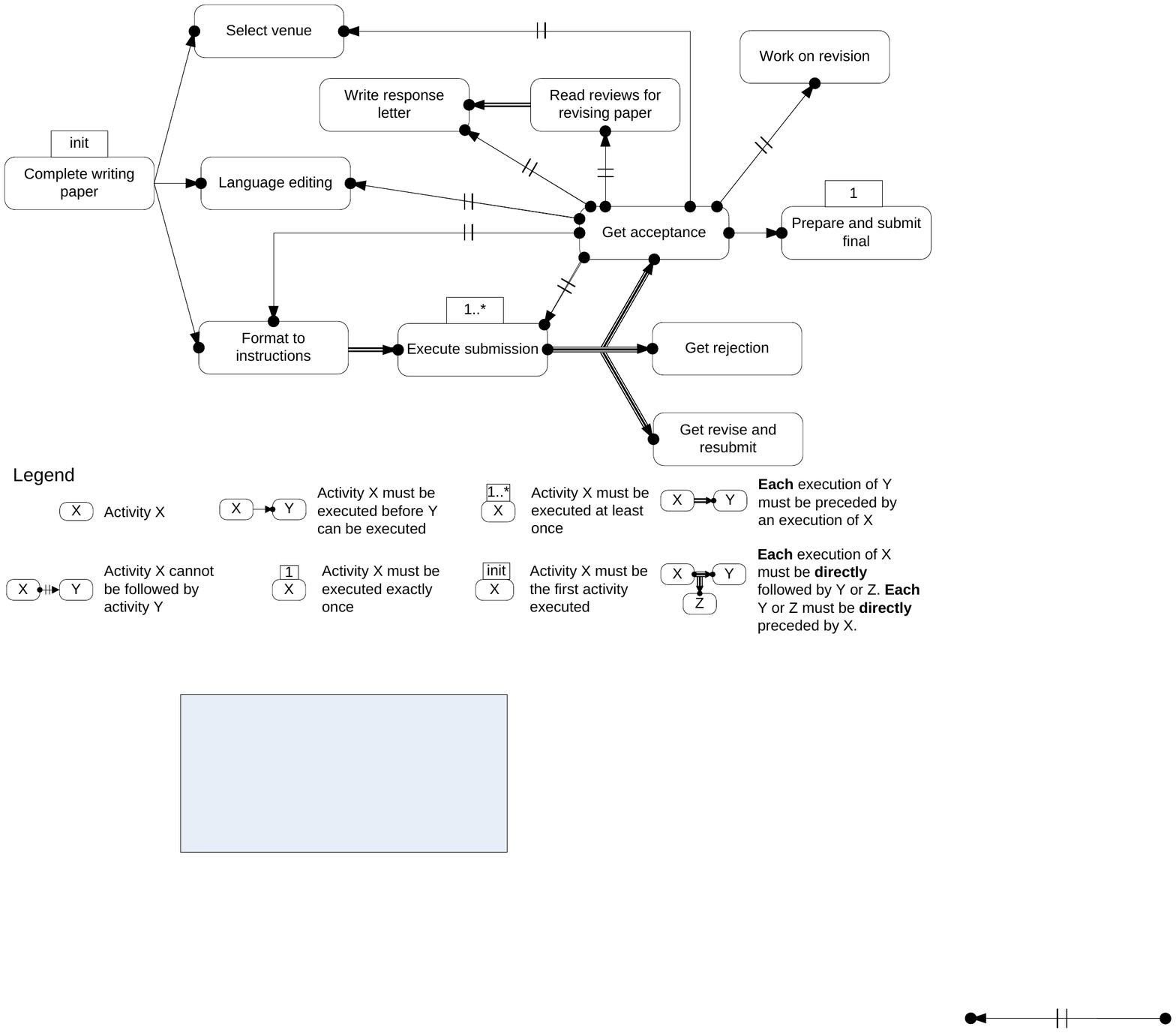}
  \caption{Example of a flat model}
  \label{fig:example_flat}
\end{center}
\end{figure}

\begin{figure}[htp]
\begin{center}
  \includegraphics[width=\textwidth]{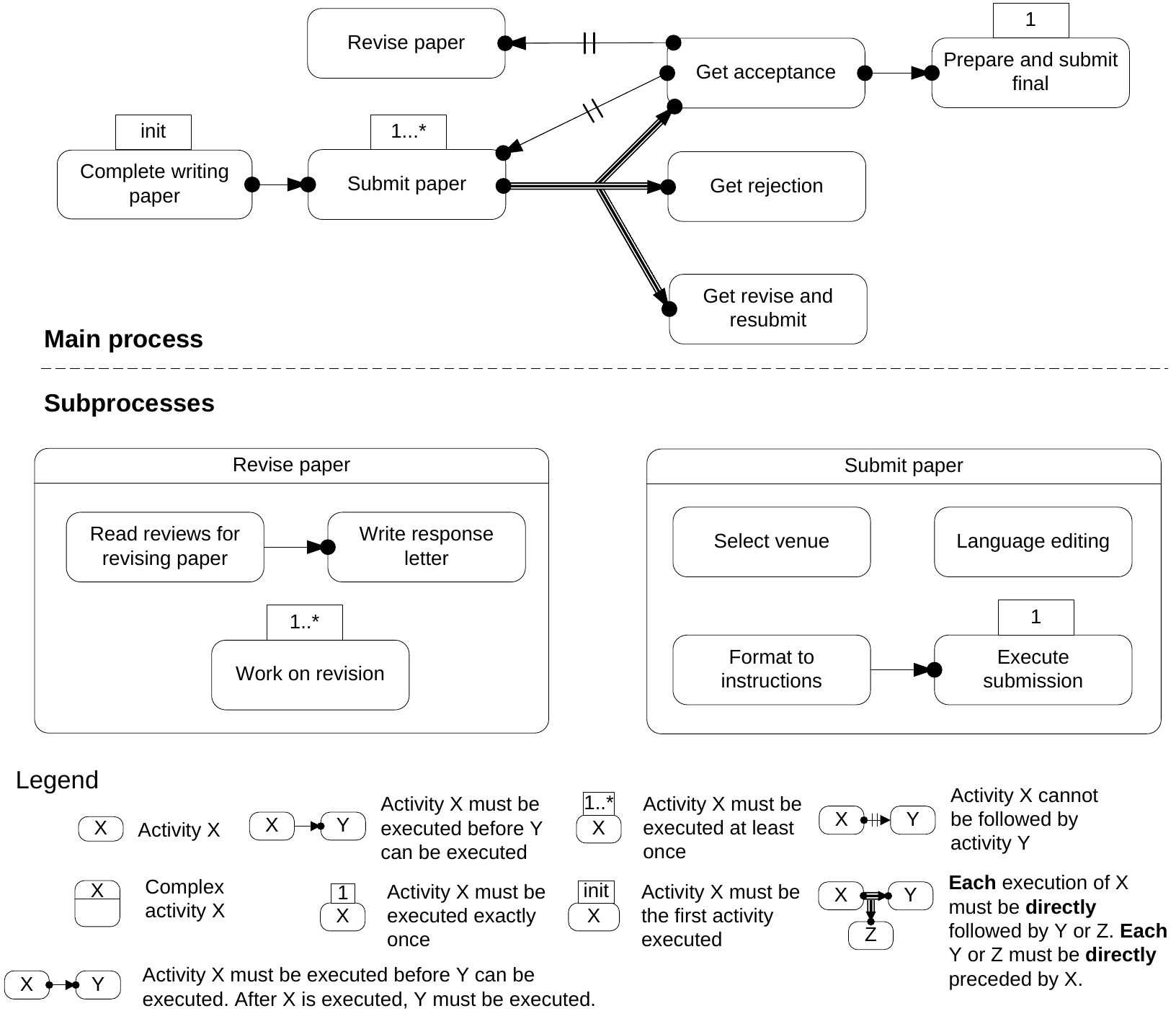}
  \caption{Example of a hierarchical model}
  \label{fig:example_hierarchical}
\end{center}
\end{figure}

To illustrate and discuss the implications of hierarchy on declarative process
models, we make use of a running example. We chose the business process of
writing a scientific paper and created two business process \textit{models}
describing the process. In \figurename~\ref{fig:example_flat} the process is
modeled without hierarchy, whereas in \figurename~\ref{fig:example_hierarchical}
hierarchical structures are used. Due to space restrictions, the examples are not
described in detail, but will be used in the following for illustration purposes.

\subsection{Preconditions for Using Sub-Processes}
\label{sec:preconditions}
While for imperative models any Single-Entry-Single-Exit fragment can be
extracted to a sub-process~\cite{Web+11}, \cite{Web+08}, in declarative models
the structure is not informative enough. Rather, two main conditions should hold for
the introduction of sub-processes. First, the activities in a sub-process should
relate to a certain intention~\cite{SoRo05} to be fulfilled.\footnote{Formalizing
intentions in declarative process models is out of scope for this work and will
be left for future research.} For instance, in
\figurename~\ref{fig:example_hierarchical}, \textit{Read reviews for revising
paper}, \textit{Write response letter} and \textit{Work on revision} all serve
the purpose of revising a paper. Once the sub-process of \textit{Revise paper} is
completed, it is clear that the paper has been revised. On a higher abstraction
level it may not make a difference, e.g., how many times \textit{Work on
revision} has been executed or whether the reviews have been read. But knowing
the paper has been revised is substantial for the continuation of the process.
This information is not available in the flat model (and it only exists in the
mind of the human who executes the process). Second, the activities included in a
sub-process should be such that they can be executed in isolation from the
top-level process. This is due to the local nature of the constraints within the
sub-process, and the lack of communication with the parent process, as discussed
in Section~\ref{sec:semantics}. In other words, a sub-process cannot include any
activity that has constraints specifically relating that activity to activities
outside the sub-process. Still, if all the activities considered for inclusion in
a sub-process share a common constraint with some other activity, then this
constraint holds for the entire sub-process. In the flat model (cf.
\figurename~\ref{fig:example_flat}), activities \textit{Read reviews for revising
paper}, \textit{Write response letter} and \textit{Work on revision} all have a
constraint restricting them from following \textit{Get acceptance}. In the
hierarchical model (cf. \figurename~\ref{fig:example_hierarchical}), these
constraints are \textit{aggregated} to one constraint related to the top-level
complex activity of \textit{Revise paper}. As the constraints are aggregated so a
single constraint, we refer this to as \textit{aggregation of constraints}.

\subsection{Enhanced Expressiveness}
\label{sec:enhanced_expressiveness}
For \textit{imperative} process models, hierarchical decomposition is viewed as a
structural measure that may impact model understandability~\cite{Zug+11}, but
does not influence semantics. In declarative process models, however, hierarchy
also has implications on semantics. More precisely, hierarchy enhances the
expressiveness of a declarative modeling language. The key observation is that
by specifying constraints that refer to complex activities it is possible to
restrict the life-cycle of a sub-process. A constraint that refers to a
complex activity thereby not only influences the complex activity, but also all
activities contained therein.

This, in turn leads to two effects. First, constraints can be specified that
apply for a set of activities (cf. aggregation of constraints in
Section~\ref{sec:preconditions}). Second, the specification of constraints, that
apply in a certain context only, is supported. Consider for instance \textit{Work
on revision} and \textit{Revise paper} in
\figurename~\ref{fig:example_hierarchical}. \textit{Work on revision} is
mandatory within the context of \textit{Revise paper}. Hence, \textit{Work on
revision} must be executed at least once whenever \textit{Revise paper} is
executed, but it might not be executed at all (if \textit{Revise paper} is not
executed).

\begin{figure}[htp]
\begin{center}
  \includegraphics[width=\textwidth]{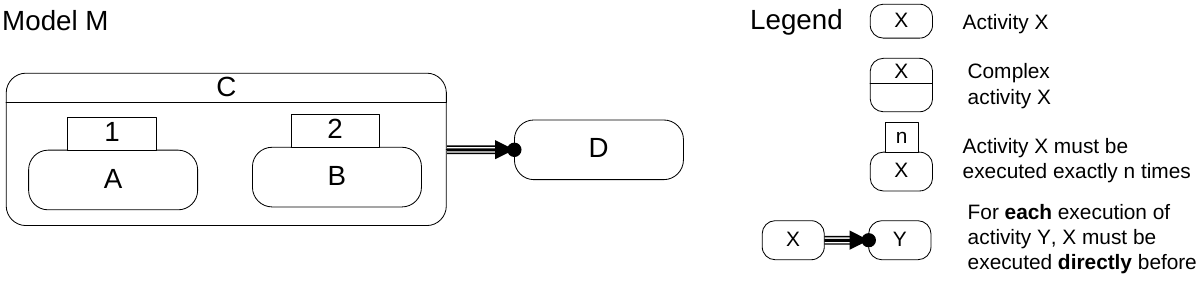}
  \caption{Enhanced expressiveness}
  \label{fig:expressiveness}
\end{center}
\end{figure}

To illustrate how these two effects enhance expressiveness, consider model
\textit{M} in \figurename~\ref{fig:expressiveness}, which solely uses constraints
defined in~\cite{Pesi08}. The chained precedence constraint between \textit{C}
and \textit{D} specifies that for each execution of \textit{D}, sub-process
\textit{C} has to be executed directly before. When executing sub-process
\textit{C}, in turn, \textit{A} has to be executed exactly once and \textit{B}
has to be executed exactly twice (in any order). Hence, the constraint between
\textit{C} and \textit{D} actually refers to a \textit{set of activities}. For
each execution of \textit{D}, \textit{A} has to be executed exactly once and
\textit{B} has to be executed exactly twice. In other words, constraints on
\textit{A} and \textit{B} are only valid in the \textit{context of C}. Such
behavior cannot be modeled without hierarchy, using the same set of constraints.

%

\subsection{Impact on Adaptation}
\label{sec:structural_changes}
Constructing hierarchical models supports top-down analysis, i.e., creating the
top-level model first and further refining complex activities thereafter. While
this seems like a natural way of dealing with complexity, in some cases, it is
desirable to transform a flat model to a hierarchical one. In the following we
will argue why refactoring~\cite{Web+11}, i.e., changing hierarchical structures
in a control-flow preserving way, is only possible under certain conditions for
declarative process models. Refactoring requires that \textit{any} hierarchical
model can be translated into a model without hierarchy, but the same control-flow
behavior (and vice versa). As discussed, expressiveness is enhanced by hierarchy.
In other words, there exists control flow behavior that can be expressed in an
hierarchical model, but not in a model without hierarchy---cf.
\figurename~\ref{fig:expressiveness} for an example. Hence, only those
hierarchical models that do not make use of the enhanced expressiveness can be
refactored.

\section{Model Understandability}
\label{sec:understandability}
So far we discussed that hierarchy in declarative process models is not just a
question of structure, but also affects semantics. In the following, we will
describe how these effects impact the understandability of a declarative
process model.

\subsection{Framework for Assessing Understandability}
\label{sec:theory}
The influence of hierarchy on model understandability has been investigated in a
number of different modeling languages, such as ER-Models~\cite{Mood04},
imperative business process models~\cite{Rej+11} and UML
Statecharts~\cite{Cru+05a} (for an overview see~\cite{Zug+11}).
While reported results do not entirely clarify when and how understandability is
affected, a trade-off between (sub)model size and degree of hierarchy can be
observed. For instance, in small models hierarchy may have no~\cite{Cru+05a} or
even a negative impact~\cite{Cru+08}, while for large models a positive
influence could be observed~\cite{Rej+11}.

\begin{figure}[htp]
\begin{center}
  \includegraphics[width=\textwidth]{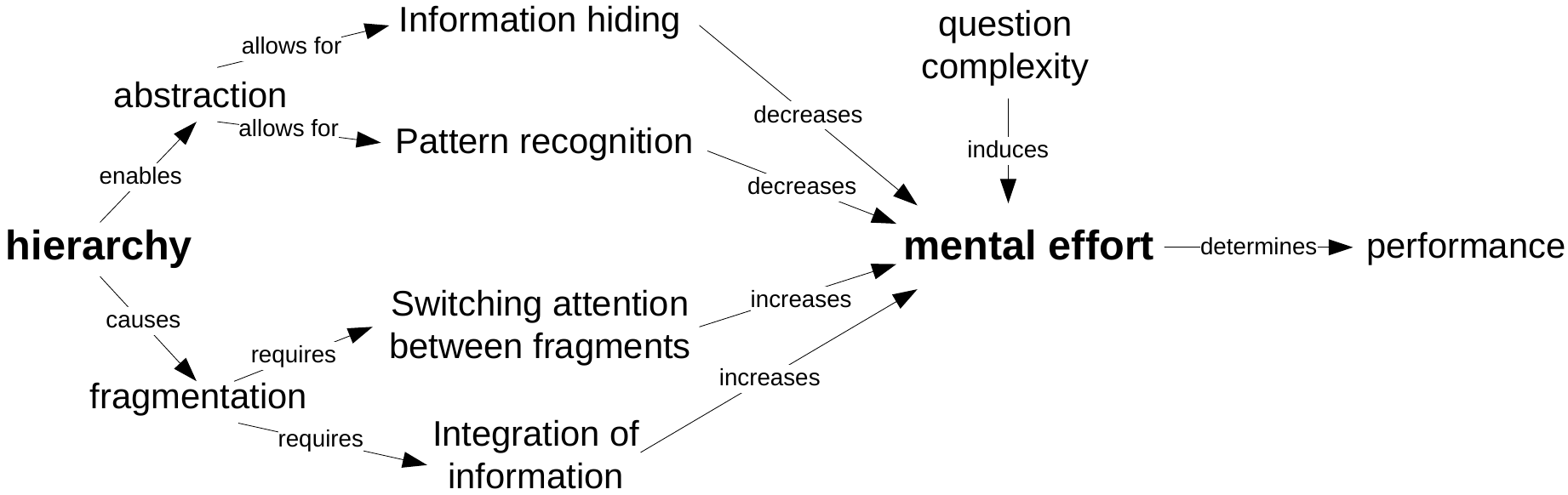}
  \caption{Framework for assessing hierarchy, adapted from~\cite{Zug+11}}
  \label{fig:framework}
\end{center}
\end{figure}

In~\cite{Zug+11}, we introduced a cognitive-psychology-based theory describing
when and why hierarchy has an impact on understandability (for a introduction to
cognitive psychology in business process modeling we refer to~\cite{Zug+11a}).
In this work we present an enhanced version that is still generic but also
takes into account the idiosyncrasies of hierarchy in declarative process models.
The central concept of the framework is \textit{mental effort}~\cite{Swel88},
i.e., the mental resources required to solve a problem. In the context of this
work, solving a problem refers to understanding the semantics of a declarative process
model, i.e., answering questions about a model. According to the framework,
hierarchy is the source of two opposing forces influencing this problem solving
process. Positively, \textit{abstraction} decreases mental effort by
\textit{hiding information} and supporting the \textit{recognition of patterns}.
Negatively, \textit{fragmentation} increases mental effort by forcing the reader
to \textit{switch attention between fragments} and \textit{integrating}
information from fragments.

\paragraph {Abstraction.} Hierarchy allows to aggregate model information by
hiding the internals of a sub-process using a complex activity. Thereby,
irrelevant information can be hidden from the reader, leading to decreased mental
effort, as argued in~\cite{Mood04}. From the perspective of cognitive psychology,
this phenomenon can be explained by the concept of \textit{attention
management}~\cite{LaSi87}. During the problem solving process, i.e., answering a
question about a model, attention needs to be guided to certain parts of a model.
For instance, when checking whether a certain execution trace is supported by a
process model, activities that are not contained in the trace are irrelevant for
answering the question. Here, abstraction allows removing this irrelevant
information, in turn supporting the attention management system and thus reducing
mental effort. To illustrate this effect for declarative process models, consider
the process model shown in \figurename~\ref{fig:example_hierarchical}. For
answering the question, whether \textit{Get acceptance} can be executed after
\textit{Complete writing paper} it is sufficient to look at activities
\textit{Complete writing paper}, \textit{Submit paper} and \textit{Get
acceptance}. In other words, hierarchy helps to abstract from all activities
contained in \textit{Submit paper}, making the question easier to answer.

Besides reducing mental effort by improving attention management, abstraction
presumably supports the identification of higher level patterns. It is known that
the human's perceptual system requires little mental effort for recognizing
certain patterns~\cite{LaSi87}, \cite{ScRo96}, e.g., recognizing a well-known
person does not require thinking, rather this information can be directly \textit{perceived}.
Similarly, in process models, by abstracting and thereby aggregating information,
presumably information can be easier perceived. Consider for example the process
models depicted in \figurename~\ref{fig:example_flat} and
\figurename~\ref{fig:example_hierarchical}. The models are (almost) information
equivalent, still we argue that for the model with sub-processes the overall
structure and intention of the process is easier to grasp. By introducing complex activities, it
is easier to see that the process is about iteratively reworking a paper
until it gets accepted. For the sibling-model in
\figurename~\ref{fig:example_flat}, however, the reader first has to mentally
group together activities before the overall intention of the process becomes
clear.

\paragraph {Fragmentation.} Empirical evidence shows that the influence of
hierarchy can range from positive over neutral to negative (cf. \cite{Rej+11},
\cite{Cru+08}, \cite{Cru+05a}, \cite{Mood04}). To explain the negative influence, we refer to the \textit{fragmentation} of the model. When extracting a
sub-process, modeling elements are removed from the parent model and placed
within the sub-process. When answering a question that also refers to the
content of a sub-process, the reader has to \textit{switch attention}
between the parent model and the sub-process. In addition, the reader has to
\textit{mentally integrate} the sub-process into the parent model, i.e.,
interpreting constraints in the context of the parent process. From the
perspective of cognitive psychology, these phenomena are known to increase mental effort and
referred to as \textit{split-attention effect}~\cite{SwCh94}. To exemplify this
effect, consider the process model in
\figurename~\ref{fig:example_hierarchical}. To determine how often activity
\textit{Execute submission} must be executed, it is required to look at
activity \textit{Submit paper} too, as \textit{Execute submission} is contained
therein. In other words, the reader has to split attention between these two
activities. In addition, the reader has to integrate the execution semantics of
\textit{Submit paper} with the execution semantics of \textit{Execute
submission}. Both activities are mandatory, i.e., must be executed at least
once, hence for any execution of the overall process, \textit{Execute
submission} must be executed at least once. In other words, it is necessary to
mentally integrate the constraints restricting  the execution of \textit{Submit
paper} as well as constraints restricting the execution of \textit{Execute
submission}.

\paragraph {Interplay of Abstraction and Fragmentation.} According to
the model illustrated in \figurename~\ref{fig:framework}, a question's complexity
induces a certain mental effort, e.g., locating an activity is easier than
validating an execution trace. In addition, mental effort may be decreased by
information hiding and pattern recognition, or increased by the need to switch
between sub-processes and integrate information. Thereby, abstraction as well as
fragmentation occur at the same time. A model without sub-processes apparently
cannot benefit from abstraction, neither is it impacted by fragmentation. By
introducing hierarchy, i.e., creating sub-processes, \textit{both} abstraction
and fragmentation are stimulated. Whether the introduction of a new sub-process
influences understandability positively or negatively then depends on whether the
influence of abstraction or fragmentation predominates. For instance, when
introducing hierarchy in a small process model, not too much influence of
abstraction can be expected, as the model is small anyway. However, fragmentation
will appear, regardless of model size. In other words, hierarchy will most likely
show a negative influence or at best no influence for small models
(cf.~\cite{Cru+08}, \cite{Cru+09}).

\subsection{Impact of Idiosyncrasies on Understandability}
\label{sec:impact_of_idiosyncrasies}
In Section~\ref{sec:using_hierarchy}, we have shown that hierarchy enhances
expressiveness and allows to aggregate constraints. In the following, we will
discuss the impact of these two phenomena on understandability.

\paragraph {Enhanced Expressiveness and Complex Mental Integration.} As argued,
hierarchy provides enhanced expressiveness. However, this also comes at a price,
as the constraint that is referring to a sub-process has to be integrated with
the semantics of the constraints \textit{within} the sub-process. To illustrate
such integrations, consider the process model in
\figurename~\ref{fig:example_hierarchical}. Activity \textit{Work on revision}
has to be executed at least once, i.e., is mandatory. However, this activity is
contained in complex activity \textit{Revise paper}, which is optional. In other
words, \textit{Work on revision} is mandatory for \textit{Revise paper}, which is
optional for the main process. Consequently, also \textit{Work on revision} is
optional for the main process.

Such mental integrations can be found in any hierarchical conceptual model. For
instance, in an imperative process model, mental integration refers to
transferring the token from the parent process to the start event of the
sub-process. As argued, however, integrations are particularly complex in
declarative process models. Hence, it can be expected that a strong influence on
the understandability can be observed.

\paragraph {Aggregation of Constraints.} As discussed in
Section~\ref{sec:enhanced_expressiveness}, hierarchy allows to aggregate and
thus reduce the number of constraints. In the context of the proposed
framework, we can identify three forces. Positively, aggregation reduces the number of
constraints, hence hiding information. In addition, a reduced number of
constraints fosters the the layout of the process model. This, in turn, supports
the recognition of patterns, i.e., making the model easier to understand.
Negatively, complex mental integration operations, as discussed before, may
diminish the described gains. Whether positive or negative influences
predominate will have to be investigated empirically, as discussed in the
following.

\subsection{Discussion}
\label{sec:discussion}
So far we argued that hierarchy in declarative process models can be attributed
to increases as well as decreases in understandability. In the following, we will
discuss the impact of the identified influences. Positively, we see a big
potential for hierarchy in declarative process models. In an imperative process
model, control flow is modeled explicitly. Hence, process models are usually
structured according to their control flow. Such a strategy is in general not
possible for a declarative process model, as constraints do not necessarily
prescribe sequential information. Sub-processes, however, allow to group
activities and thereby to introduce structure to the model. Sub-processes,
however, allow to group activities by a mutual intention they serve and thereby
to introduce structure to the model and add higher-level information. As argued
in our framework, this allows \textit{recognizing patterns} and makes it easier
to grasp the intention of a business process (cf.
\figurename~\ref{fig:example_flat} and
\figurename~\ref{fig:example_hierarchical}). Also the ability of sub-processes to
\textit{hide information}, i.e., activities and constraints, can be expected to
contribute to the understandability of models. It is assumed that several
interconnected constraints quickly become challenging for the human
mind~\cite{ZuPWxx}, \cite{Pesi08}, \cite{Web+09b}. Hence, hiding information
can be expected to be especially beneficial in declarative process models.

On the other hand, as argued in Section~\ref{sec:impact_of_idiosyncrasies}, we
assume that the integration of constraints poses a significant challenge for the
reader. In particular, it is not clear yet whether an average process modeler is
able to efficiently perform such mental integrations. This is, however, necessary
for the meaningful application of enhanced expressiveness by hierarchy. If
efficient mental integration was not possible, enhanced expressiveness would be
rendered useless as resulting models would be hardly understandable.

The presented framework can be seen as a first step towards a systematic
assessment of the impact of hierarchy on understandability in declarative process models.
Even though it is based upon well-established concepts from cognitive
psychology, the claims still have to be empirically challenged. In particular, we
postulated that the integration of constraints poses a significant, but
manageable challenge for the reader. Similarly, we assume that \textit{large}
declarative process models tend to be too complex for humans to deal with
(cf.~\cite{Pesi08}). To corroborate the postulated claims, we are
currently planning a thorough empirical investigation, cf.
Section~\ref{sec:summary}. Therein, we plan to assess understandability in two
ways. First, we assume that the easier the understanding of a model is, the less
mental effort is required. Hence, we will use Likert-scales to assess mental
effort (for the reliability of such measures see~\cite{Paa+03}). Second, and
similar to~\cite{Rej+11}, \cite{Cru+08}, \cite{Cru+09}, we will measure the
ratio of correctly answered questions as well as the required duration. Apparently,
hierarchical models will not be allowed to use constructs that go beyond the
expressiveness for non-hierarchical models. Otherwise, information equivalency
is not given anymore, imparing the experimental setup. Due to this reason, we
plan to conduct experiments for such models first and hope to be able to
extrapolate findings to expressiveness-enhanced models.

\section{Related Work}
\label{sec:related_work}
In this work we discussed characteristics of hierarchy in declarative process
models and the impact on understandability. The impact of hierarchy on
understandability has been studied in various conceptual modeling languages, such
as imperative business process models~\cite{Rej+11}, ER
diagrams~\cite{Mood04}, \cite{ShDM04} and UML statechart
diagrams~\cite{Cru+08}, \cite{Cru+09}, \cite{Cru+07} (an overview is presented
in~\cite{Zug+11}). Still, none of these works deals with the impact of hierarchy on
understandability in declarative process models. The understandability of
declarative process models in general has been investigated in the work of Zugal
et al.~\cite{ZuPWxx}, \cite{ZuPW11}, \cite{Zug+11b}, however, in contrast to
this work, hierarchy is not discussed. With respect to understandability of process models in general,
work dealing with the understandability of imperative business process models is
related. In~\cite{MeRA10} modeling guidelines are presented that target to
improve the understandability of imperative process models. The understandability
of imperative process models is investigated empirically in~\cite{ReMe}.
Finally, in~\cite{Pesi08} the technical aspects of declarative business
process models, such as the definition of modeling languages or verification of
models is investigated. In contrast to this work, understandability aspects are
neglected and the unique semantics and expressiveness enabled by sub-processes is
not elaborated.
\section{Summary and Outlook}
\label{sec:summary}
In this work we examined hierarchy in declarative business process models. After
elaborating on the semantics, we discussed the usage and peculiarities of
hierarchy. In particular, we showed that hierarchy enhances expressiveness, but
cannot be used arbitrarily to any model fragment. Subsequently, we discussed
implications on the understandability of declarative process models. Thereby, we
built upon previous work and proposed a cognitive-theory based framework to
systematically assess the impact of hierarchy on understandability in declarative
process models. In general it can be said that hierarchy should be handled with
care. On the one hand, information hiding and increased pattern recognition
promise gains in terms of understandability. On the other hand, the integration
of constraints presumably poses a significant challenge for the reader. In
addition, switching between sub-processes may compromise the understandability of
respective models. We acknowledge that, even though the framework is based on
well-established concepts from cognitive psychology, an empirical validation
still has to be conducted.

In this sense, our next steps clearly focus on empirical validation. In
particular, two main research directions are envisioned. First, we will
investigate whether information hiding lowers cognitive load and hence improves
understandability. Second, the integration of constraints and the required mental
effort will be scrutinized.

\bibliographystyle{splncs}
\bibliography{literature}

\end{document}